\documentclass[twocolumn,showpacs,aps]{revtex4}

\usepackage{graphicx}
\usepackage{latexsym}
\usepackage{amssymb}
\usepackage{color}
\usepackage[center]{subfigure}
\input psfig.sty

\begin{document}

\newcommand{\bq}{\begin{equation}}
\newcommand{\eq}{\end{equation}}
\newcommand{\bqn}{\begin{eqnarray}}
\newcommand{\eqn}{\end{eqnarray}}
\newcommand{\nb}{\nonumber}
\newcommand{\lb}{\label}

\title{Shear-free Dust Solution in General Covariant Ho\v{r}ava-Lifshitz Gravity}
 
\author{O. Goldoni $^{1}$}
\email{otaviosama@gmail.com}
\author{M.F.A. da Silva $^{1}$}
\email{mfasnic@gmail.com}
\author{R. Chan $^{2}$}
\email{chan@on.br}
\affiliation{
$^{1}$ Departamento de F\'{\i}sica Te\'orica,
Instituto de F\'{\i}sica, Universidade do Estado do Rio de Janeiro,
Rua S\~ao Francisco Xavier 524, Maracan\~a
20550-900, Rio de Janeiro, RJ, Brasil.\\
$^{2}$ Coordena\c{c}\~ao de Astronomia e Astrof\'{\i}sica, 
Observat\'orio Nacional, Rua General Jos\'e Cristino, 77, S\~ao Crist\'ov\~ao  
20921-400, Rio de Janeiro, RJ, Brazil.}

\date{\today}

\begin{abstract}
In this paper, we have studied non stationary dust spherically symmetric
spacetime, in general covariant theory ($U(1)$ extension) of the Ho\v{r}ava-Lifshitz
gravity with the minimally coupling and non-minimum coupling with matter, 
in the post-newtonian approximation (PPN) in the infrared limit. 
The Newtonian prepotential $\varphi$ was assumed null. 
The aim of this work is to know if we can have the same spacetime, as we know in the 
General Relativity Theory (GRT), in Ho\v{r}ava-Lifshitz Theory (HLT) in this limit.
We have shown that there is not an analogy of the dust solution in 
HLT with the minimally coupling, as in GRT.
 Using non-minimum coupling with matter, we have shown that the solution
admits a process of gravitational collapse, leaving a singularity at the end.
This solution has, qualitatively, the same temporal behaviour as the dust collapse in GRT.
However, we have also found a second possible solution, representing a bounce behavior 
that is not found in GRT.
\end{abstract}


\maketitle

\section{Introduction}

The construction of a Quantum Theory of Gravitation is one of the unsolved problems of
modern physics. The biggest challenge faced by all the 
researchers, who have tried to build a new gravitation theory, is that this new theory must be
valid at all scales. Since the gravity force is a  relatively weak one,
it is not expected to be observed any of its quantum effects. For this
 reason the criteria that a new candidate quantum gravity theory must fulfill are
 constrained by mathematical and self-consistency criteria, reproducing observable results, confirming General Relativity Theory, and finally, making non-trivial predictions that may possibly be tested.

Ho\v{r}ava-Lifshitz gravity theory (HLT) \cite{Horava}
\cite{Lifshitz} has drawn a lot of interest
 primarily due to its properties to solve the perturbative non-renormalizability
 of quantized standard Einstein gravity at the expense of  breaking
 relativistic invariance at very high energies, while restoring it at low
 energies. After that,  further deep
 connections of Ho\v{r}ava-Lifshitz formalism to condensed matter physics were
 discovered within the framework of gauge/gravity duality (holography).

 Unfortunately, the original Ho\v{r}ava formulation was contaminated by a number 
of serious problems which has caused the development of various non-trivial
 modifications. One of the promising latter modifications is the one
without the so called projectability condition and with an additional Abelian
gauge symmetry, which in particular solves the problem with the unphysical
scalar graviton. Due to the extensive works in this area, we suggest to
the reader the references \cite{Visser}-\cite{Lin:2012bs}.

In order to establish the physical relevance of any modification of
 Ho\v{r}ava-Lifshitz gravity it is very important to check that the latter
 reproduces the known physically feasible properties of standard Einstein
 General Relativity (EGR) at low energies. Within this condition, we 
have checked whether the non-projectable version of Ho\v{r}ava-Lifshitz gravity 
with the extra U(1) gauge symmetry contains in the low energy limit solutions of 
Vaidya-type \cite{Goldoni2014}. The answer was negative 
and, therefore, led us to the important conclusion that in order 
to establish consistency with EGR at low energies the
gauge field associated with the extra U(1) gauge symmetry of the enlarged
non-projectable Ho\v{r}ava-Lifshitz gravity should have some interaction with 
the pure radiation matter generating the Vaidya spacetime geometry. 

Lin et al. (2014) \cite{Lin2014},
have proposed a universal coupling between the gravity
and matter in the framework of the Ho\v{r}ava-Lifshitz theory of gravity
with an extra U(1) symmetry for both the projectable and
non-projectable cases. Then, using this universal coupling they have 
studied the PPN  approximations and they have obtained the 
PPN parameters in terms of the coupling constants of the theory. 

Goldoni et al. (2014) \cite{Goldoni2014}, using Lin 
at al. (2014) \cite{Lin2014} approximation, have shown that there is not an 
analogy of the Vaidya's solution in 
Ho\v{r}ava-Lifshitz Theory (HLT) without projectability, as we know
 in General Relativity Theory (GRT).

In another recent paper, Goldoni et al. (2015) \cite{Goldoni2015}, using again Lin 
at al. (2014) \cite{Lin2014} approach, have also shown that there is not an analogy 
of the Vaidya's solution in 
Ho\v{r}ava-Lifshitz Theory (HLT) with the minimally coupling and without projectability, again as we know in GRT.

The goal of this present paper is to study the behavior of a dust fluid solution of GRT in HLT for the infrared limit. 

We want to know if we can have the same spacetime, as we know in 
GRT, in HLT in this limit.

 Using again the results of Lin et al. (2014) \cite{Lin2014}
 we have studied the spherically 
symmetric spacetime filled by a dust fluid,  in general covariant theory ($U(1)$ extension) of
Ho\v{r}ava-Lifshitz gravity with a minimally coupling \cite{Lin2014}, in the
 PPN approximation and in the infrared 
limit.  We will analyze if a solution like this one can be described in the general
covariant HLT of gravity \cite{ZWWS,ZSWW}.
In Section II we present a brief introduction to HLT with the minimally
 coupling \cite{Lin2014} considered here and present the field equations of HLT modified.
In Section III we show the dust solution in the infrared limit.
In Section IV we discuss the results. Finally, in
 the Appendix A we present some quantities necessary in HLT field equations
 with minimally coupling \cite{Lin2014}.

\section{General Covariant Ho\v{r}ava-Lifshitz Gravity with Coupling with Matter}

In this section, we shall give a very brief introduction to the general 
covariant HLT gravity with the minimally coupling. For detail, we refer readers to \cite{ZWWS,ZSWW,Lin2014}.

The Arnowitt-Deser-Misner (ADM) form is given by \cite{ADM},
\bqn
ds^{2} &=& - N^{2}dt^{2} + g_{ij}\left(dx^{i} + N^{i}dt\right)
\left(dx^{j} + N^{j}dt\right), \nb\\
& & ~~~~~~~~~~~~~~~~~~~~~~~~~~~~~~  (i, \; j = 1, 2, 3),
\lb{ds2}
\eqn
where the non-projectability condition imposes that $N=N(t,x^i)$.

In the work of Lin et al. (2014) \cite{Lin2014} it is proposed that, in the IR limit, it is possible to have matter fields universally
couple to the ADM components through the transformations
\bqn
\lb{eq8-1}
& & \tilde{N} = F N,\;\;\;
\tilde{N}^i = N^i + Ng^{ij} \nabla_j\varphi,\nb\\
&& 
\tilde{g}_{ij} = \Omega^2g_{ij},
\eqn
with
\bqn
\lb{eq8-1a}
& & F = 1 - a_1\sigma, \;\;\;
 \Omega = 1 - a_2\sigma,
\eqn
where
\bqn
\sigma &\equiv &\frac{A - {\cal{A}}}{N},\nb\\
{\cal{A}} &\equiv& - \dot{\varphi}  + N^i\nabla_i\varphi
+\frac{1}{2}N\left(\nabla^i\varphi\right)\left(\nabla_i\varphi\right),\nb\\
\eqn
and where $A$ and $\varphi$ 
are the gauge field and the Newtonian prepotential,
 respectively, 
and $a_1$ and $a_2$ are two arbitrary coupling constants. Note that by
setting the first terms in $F$ and $\Omega$ to unity, we have used the
freedom to rescale the units of time and space. We also have

\bq
\lb{eq8-2}
\tilde{N}_i =\Omega^2\left(N_i + N\nabla_i\varphi\right),\;\;\;
\tilde{g}^{ij} = \Omega^{-2}g^{ij}.
\eq

Considering the exposed before, the matter action can be written as

\bqn
\lb{eu7}
S_{m} &=& \int{dtd^3x \tilde{N}\sqrt{\tilde{g}}\;  \tilde{\cal{L}}_{m}\left(\tilde{N}, \tilde{N}_i, \tilde{g}_{ij}; \psi_{n}\right)},
\eqn
where $\psi_n$ collectively  stands for matter fields. One can then define the 
matter stress-energy in the ADM decomposition, with the minimally coupling. The different components are given by (for the details see \cite{Lin2014})

\bqn
\lb{Tab}
\rho_H (EGR) & = & T_{\mu\nu}n^\mu n^\nu \equiv J^t =  -\frac{\delta(\tilde{N}\tilde{\cal{L}}_{m})}{\delta(\tilde{N})}\nb\\
S^i (EGR) & = & -T_{\mu\nu} h^{(4)i\mu}n^\nu \equiv J^i =  \frac{\delta(\tilde{N}\tilde{\cal{L}}_{m})}{\delta(\tilde{N}_i)}\nb\\
S^{ij} (EGR) & = & T_{\mu\nu}h^{(4)i\mu}h^{(4)j\nu} \equiv \nb\ \\
& & \tau^{ij} = \frac{2}{\tilde{N}\sqrt{\tilde{g}}}\frac{\delta(\tilde{N}\sqrt{\tilde{g}}\tilde{\cal{L}}_{m})}{\delta(\tilde{g}_{ij})},
\eqn
where ${h^{(4)}}_{\mu\nu}$ is the projection operator, defined as ${h^{(4)}}_{\mu\nu}\equiv {g^{(4)}}_{\mu\nu}+n_\mu n_\nu$ and $n^\mu$ is the normal vector to the hypersurface $t=$ constant, defined as $n^\mu = \frac{1}{N}(-1,N^i)$. Before, the prime and dot denotes the partial differentiation in relation to the
coordinate $r$ and $t$, respectively.

Thus, the total action of the theory can be written as,
\bqn 
\lb{TA}
S &=& \zeta^2\int dt d^{3}x  \sqrt{g}N \Big({\cal{L}}_{K} -
{\cal{L}}_{{V}} +  {\cal{L}}_{{A}}+ {\cal{L}}_{{\varphi}}  + {\cal{L}}_{S}+\nb\\
& &\frac{1}{\zeta^2} {\cal{L}}_{M}\Big), 
\eqn
where $g={\rm det}(g_{ij})$, $N$ is given in the equation (\ref{ds2}) and
\bqn \lb{2.5}
{\cal{L}}_{K} &=& K_{ij}K^{ij} -   \lambda K^{2},\nb\\
{\cal{L}}_{V} &=&  \gamma_{0}\zeta^{2}  -  \Big(\beta_0  a_{i}a^{i}- \gamma_1R\Big)
+ \frac{1}{\zeta^{2}} \Big(\gamma_{2}R^{2} +  \gamma_{3}  R_{ij}R^{ij}\Big)\nb\\
& & + \frac{1}{\zeta^{2}}\Bigg[\beta_{1} \left(a_{i}a^{i}\right)^{2} + \beta_{2} \left(a^{i}_{\;\;i}\right)^{2}
+ \beta_{3} \left(a_{i}a^{i}\right)a^{j}_{\;\;j} \nb\\
& & + \beta_{4} a^{ij}a_{ij} + \beta_{5}
\left(a_{i}a^{i}\right)R + \beta_{6} a_{i}a_{j}R^{ij} + \beta_{7} Ra^{i}_{\;\;i}\Bigg]\nb\\
& & +  \frac{1}{\zeta^{4}}\Bigg[\gamma_{5}C_{ij}C^{ij}  + \beta_{8} \left(\Delta{a^{i}}\right)^{2}\Bigg],\nb\\
{\cal{L}}_{A} &=&-\frac{AR}{N}, \nb\\
{\cal{L}}_{\varphi} &=&  \varphi{\cal{G}}^{ij}\big(2K_{ij}+\nabla_i\nabla_j\varphi+a_i\nabla_j\varphi\big)\nb\\
& & +(1-\lambda)\Big[\big(\Delta\varphi+a_i\nabla^i\varphi\big)^2  
+2\big(\Delta\varphi+a_i\nabla^i\varphi\big)K\Big]\nb\\
& & +\frac{1}{3}\hat{\cal G}^{ijlk}\Big[4\left(\nabla_{i}\nabla_{j}\varphi\right) a_{(k}\nabla_{l)}\varphi \nb\\
&&  ~~ + 5 \left(a_{(i}\nabla_{j)}\varphi\right) a_{(k}\nabla_{l)}\varphi + 2 \left(\nabla_{(i}\varphi\right)a_{j)(k}\nabla_{l)}\varphi \nb\\
&& + 6K_{ij} a_{(l}\nabla_{k)}\varphi \Big], \nb \\
{\cal{L}}_S &=&\sigma  (\sigma_1 a^ia_i+\sigma_2a^i_{\;\;i}),
\eqn
where
\bq
\zeta^{2} = \frac{1}{16\pi G},
\eq
where $G$ denotes the Newtonian constant, 
${\cal{L}}_M$ is the Lagrangian of matter fields,
$\hat{\cal G}^{ijlk} =  g^{il}g^{jk} - g^{ij}g^{kl}$
\cite{Lin2013}. 
Here $\Delta \equiv g^{ij}\nabla_{i}\nabla_{j}$ and all the coefficients, $ \beta_{n}$ and $\gamma_{n}$, are
dimensionless and arbitrary.  
$C_{ij}$ denotes the Cotton tensor, defined by
\bq
\lb{1.12}
C^{ij} = \frac{ {{e}}^{ikl}}{\sqrt{g}} \nabla_{k}\Big(R^{j}_{l} - \frac{1}{4}R\delta^{j}_{l}\Big),
\eq
with  $e^{123} = 1$.  Using the Bianchi identities, one can show that 
$C_{ij}C^{ij}$ can be written in terms of the five independent sixth-order 
derivative terms in the form
\bqn
\lb{1.13}
C_{ij}C^{ij}  &=& \frac{1}{2}R^{3} - \frac{5}{2}RR_{ij}R^{ij} + 3 R^{i}_{j}R^{j}_{k}R^{k}_{i}  +\frac{3}{8}R\Delta R\nb\\
& &  +
\left(\nabla_{i}R_{jk}\right) \left(\nabla^{i}R^{jk}\right) +   \nabla_{k} G^{k},
\eqn
where
\lb{1.14}
\bqn
G^{k}=\frac{1}{2} R^{jk} \nabla_j R - R_{ij} \nabla^j R^{ik}-\frac{3}{8}R\nabla^k R.
\eqn

The Ricci and Riemann tensors 
$R_{ij}$ and $R^{i}_{\;\; jkl}$  all refer to the 3-metric $g_{ij}$, with 
$R_{ij} = R^{k}_{\;\;ikj}$ and
\bqn \lb{2.6}
 R_{ijkl} &=& g_{ik}R_{jl}   +  g_{jl}R_{ik}  -  g_{jk}R_{il}  -  g_{il}R_{jk}\nb\\
 &&    - \frac{1}{2}\left(g_{ik}g_{jl} - g_{il}g_{jk}\right)R,\nb\\
K_{ij} &\equiv& \frac{1}{2N}\left(- \dot{g}_{ij} + \nabla_{i}N_{j} +
\nabla_{j}N_{i}\right),\nb\\
{\cal{G}}_{ij} &\equiv& R_{ij} - \frac{1}{2}g_{ij}R,\nb\\
a_{i} &\equiv& \frac{N_{,i}}{N},\;\;\; a_{ij} \equiv \nabla_{j} a_{i},\nb\\
\eqn
where $N_i$ is defined in the ADM form of the
metric \cite{ADM}, given by equation (\ref{ds2}).

The variations of the action $S$ (\ref{TA}) with respect to $N$ and $N^{i}$ 
give rise to the Hamiltonian and momentum constraints,
\bqn \label{hami}
{\cal{L}}_K + {\cal{L}}_V + F_V-F_\varphi-F_\lambda+{\cal{H}}_S= 8 \pi G J^t,\;\;
\eqn
\bqn \label{mom}
&& M_S^i+\nabla_j \bigg\{\pi^{ij} -(1-\lambda)g^{ij}\big(\nabla^2\varphi+a_k\nabla^k\varphi\big)\nb\\
&& ~~~~~~~~~~~~~~~ - \varphi {\cal{G}}^{ij} - \hat{{\cal{G}}}^{ijkl} a_l \nabla_k \varphi\bigg\} = 8\pi G J^i, ~~~~
\lb{jui}
\eqn
where
\bqn
{\cal H}_S&=&\frac{2\sigma_1}{N}\nabla_i\left[a^i\left(A-{\cal A}\right)\right]-\frac{\sigma_2}{N}\nabla^2\left(A-{\cal A}\right)\nb\\          
&&+\frac{1}{2} a_S \nabla_j\varphi\nabla^j\varphi,\nb\\       
M_S^i&=&-\frac{1}{2}a_S\nabla^i\varphi, \nb\\                                    
J^i &=& -N \frac{\delta {\cal{L}}_M}{\delta N_i},\;\;
J^t = 2 \frac{\delta (N {\cal{L}}_M)}{\delta N},\nb\\
\pi^{ij}&=&-K^{ij}+\lambda K g^{ij},
\eqn
with
\bqn
a_S=\sigma_1a_ia^i+\sigma_2a^i_i,
\eqn
and  $F_V$, $F_{\varphi}$ and $F_\lambda$ are given in the Appendix A.

Variations of $S$ with respect to $\varphi$ and $A$ yield, respectively,
\bqn \label{phi}
&& \frac{1}{2} {\cal{G}}^{ij} ( 2K_{ij} + \nabla_i\nabla_j\varphi  +a_{(i}\nabla_{j)}\varphi)\nb\\
&& + \frac{1}{2N} \bigg\{ {\cal{G}}^{ij} \nabla_j\nabla_i(N \varphi) - {\cal{G}}^{ij} \nabla_j ( N \varphi a_i)\bigg\}\nb\\
&& - \frac{1}{N} \hat{{\cal{G}}}^{ijkl} \bigg \{ \nabla_{(k} ( a_{l)} N K_{ij}) + \frac{2}{3} \nabla_{(k} (a_{l)} N \nabla_i \nabla_j \varphi)\nb\\
&& - \frac{2}{3} \nabla_{(j} \nabla_{i)} (N a_{(l} \nabla_{k)} \varphi) + \frac{5}{3} \nabla_j (N a_i a_k \nabla_l \varphi)\nb\\
&& + \frac{2}{3} \nabla_j (N a_{ik} \nabla_l \varphi)\bigg\}+\Sigma_S \nb\\
&& + \frac{1-\lambda}{N} \bigg\{\nabla^2  \left.[N (\nabla^2 \varphi + a_k \nabla^k \varphi)\right.] \nb\\
&& - \nabla^i [N(\nabla^2 \varphi + a_k \nabla^k \varphi) a_i] \nb\\
&&+ \nabla^2 (N K) - \nabla^i ( N K a_i)\bigg \}
 = 8 \pi G J_\varphi,
\eqn
where,
\bqn \label{Sigma}
\Sigma_S & = & -\frac{1}{2N}\Bigg\{\frac{1}{\sqrt{g}}\frac{\partial}{\partial t}\left[\sqrt{g}a_S\right]\nb\\
&&-\nabla_k\left[\left(N^k+N\nabla^k\varphi\right)a_S\right]\Bigg\}
\eqn
and
\bqn \label{ja}
R-a_S= 8 \pi G J_A,
\eqn
where
\bqn
J_\varphi = -\frac{\delta {\cal{L}}_M}{\delta \varphi},\;\;\;
J_A= 2 \frac{\delta ( N {\cal{L}}_M)}{\delta A}.
\eqn

On the other hand, the variation of $S$ with respect to $g_{ij}$ yields the 
dynamical equations,
\bqn \label{dyn}
\frac{1}{\sqrt{g}N} \frac{\partial}{\partial t}\left(\sqrt{g} \pi^{ij}\right)+2(K^{ik}K^j_k-\lambda K K^{ij})\nb\\
-\frac{1}{2}g^{ij}{\cal{L}}_K+\frac{1}{N}\nabla_k (\pi^{ik}N^j+\pi^{kj}N^i-\pi^{ij}N^k)\nb\\
+F^{ij}-F^{ij}_S-\frac{1}{2}g^{ij}{\cal{L}}_S+F^{ij}_a-\frac{1}{2}g^{ij}{\cal{L}}_A+F^{ij}_\varphi\nb\\
-\frac{1}{N}(AR^{ij}+g^{ij}\nabla^2A-\nabla^j\nabla^iA)
=8\pi G \tau^{ij},\;\;\;\;\;\;
\eqn
where
\bqn
\lb{tauij}
\tau^{ij}&=&\frac{2}{\sqrt{g}N} \frac{\delta(\sqrt{g}N{\cal{L}}_M)}{\delta g_{ij}}, \nb\\
\eqn
and $F^{ij}$, $F^{ij}_S$, $F^{ij}_a$ and $F^{ij}_\varphi$ are given in the 
Appendix A.

From reference \cite{Lin2014}, we have that
\bqn
\lb{eq12}
\tilde{N} &=&\tilde{N}(N, N_i, g_{ij}, A, \varphi),\nb\\
\tilde{N}_i &=&\tilde{N}_i(N, N_i, g_{ij}, A, \varphi),\nb\\
\tilde{g}_{ij} &=&\tilde{g}_{ij}(N, N_i, g_{ij}, A, \varphi).
\eqn
Thus,
\bqn                                                                           
\lb{eq14}                                                                      
J^{t} &=& 2\Omega^{3}\Bigg\{- \rho  \frac{\delta\tilde{N}}{\delta{N}}+  \frac{\delta\tilde{N}_i}{\delta{N}}S^i   + \frac{1}{2}\tilde{N}  \frac{\delta\tilde{g}_{ij}}{\delta{N}}S^{ij}\Bigg\}. ~~~~~~~~~                            \eqn                                                                         
Similarly, it can be shown that                                                
\bqn                                                                           
\lb{eq15}                                                                      
J^{i} &=& - \Omega^{3}\Bigg\{- \rho  \frac{\delta\tilde{N}}{\delta{N}_i}                                                                            +  \frac{\delta\tilde{N}_k}{\delta{N}_i}S^k   + \frac{1}{2}\tilde{N}  \frac{\delta\tilde{g}_{kl}}{\delta{N}_i}S^{kl}\Bigg\},\nb\\                             
\tau^{ij} &=& \frac{2 \Omega^{3}}{N}\Bigg\{- \rho \frac{\delta\tilde{N}}{\delta{g}_{ij}}                                         +  \frac{\delta\tilde{N}_k}{\delta{g}_{ij}}S^k   + \frac{1}{2}\tilde{N}  \frac{\delta\tilde{g}_{kl}}{\delta{g}_{ij}}S^{kl}\Bigg\},\nb\\
J_{A} &=& 2 \Omega^{3}\Bigg\{- \rho
\frac{\delta\tilde{N}}{\delta{A}}
+  \frac{\delta\tilde{N}_k}{\delta{A}}S^k   + \frac{1}{2}\tilde{N}  \frac{\delta\tilde{g}_{kl}}{\delta{A}}S^{kl}\Bigg\},\nb\\
J_{\varphi} &=& - \frac{1}{N}\Bigg\{\frac{1}{\sqrt{g}}\left(B \sqrt{g}\right)_{,t} - \nabla_{i}\Big[B\left(N^i + N \nabla^i\varphi\right)\Big]\nb\\
&& ~~~~~~~~~ - \nabla_i\left(N\Omega^5 S^i\right)\Bigg\},
\eqn
where
\bqn
\lb{eq16}
B &\equiv& - \Omega^{3}\Bigg\{a_1\rho  - \frac{2a_2\left(1- a_2\sigma\right)}{N}S^k\left(N_k + N\nabla_k\varphi\right)\nb\\
&& -  a_2\left(1- a_1\sigma\right)\left(1- a_2\sigma\right)g_{ij}S^{ij}\Bigg\}.
\eqn
Note that, if we use the projectable case of HLT then
${\cal{L}}_{S}=0$, ${M}^i_{S}=0$, ${\cal{H}}_{S}=0$, $\Sigma_{S}=0$ and $a_{S}=0$.
In order to have these quantitities in HLT, it is
only necessary the non-projectability condition \cite{Lin2014}.

\subsection{Shear-free Dust Solution with the Minimally Coupling in the Infrared Limit}

In order to be consistent with observations in the infrared limit \cite{Lin2014}, we assume that
\bq
\beta_1=\beta_2=\beta_3=\beta_4=\beta_5=\beta_6=\beta_7=\beta_8=\beta_9=0,
\eq
\bq
\gamma_0=\gamma_2=\gamma_3=\gamma_4=\gamma_5=\gamma_6=\gamma_7=\gamma_8=\gamma_9=0,
\eq
and for the PPN approximation in minimally coupling theory, we have
\bq
\beta_0=-2(\gamma_1+1),
\eq
\bq
a_1=a_2=0,
\eq
\bq
\sigma_1=0,
\eq
\bq
\sigma_2=4(1-a_1)=4.
\eq
Thus, we have the vanishing of the cosmological constant, as follows
\bq
\Lambda_g \equiv \frac{1}{2} \zeta^{2}\gamma_{0}=0.
\eq
Besides, In the infrared limit we must have
\bq
\lb{Jt}
J^t = -2 \rho.
\eq

Now we consider the case of a shear-free dust \cite{Goncalves2004}.
Thus, the metric can be rewritten as
\bq
ds^2 = -dt^2 +Y(r,t)^2 \left[ f(r)^2 dr^2 + d\Omega ^2 \right].
\lb{ds21}
\eq

Using the equations (\ref{ds2}) and (\ref{2.6}), we have
\bq
K_{rr}= - \dot Y Y f^2,
\eq
\bq
K_{\theta\theta}=- \dot Y Y,
\eq
\bq
K_{\phi\phi}=\sin(\theta)^2 K_{\theta\theta},
\eq
\bq
K=-3 \frac{\dot Y}{Y},
\eq
\bq
R_{rr}=2\frac{f'Y'Y-fY'Y+fY'^2}{fY^2},
\eq
\bq
R_{\theta\theta}=\frac{f'Y'-Y''f-f^3}{f^3Y},
\eq
\bq
R_{\phi\phi}=\sin(\theta)^2 R_{\theta\theta},
\eq
\bq
R=2\frac{2f'Y'Y-2Y''fY-f'fY'^2+f^3Y^2}{f^3Y^4},
\eq
\bq
{\cal{L}}_K=\frac{3\dot Y^2(1-3\lambda)}{Y^2},
\eq
\bq
{\cal{L}}_V=2\gamma_1\frac{2f'Y'Y-2fY''Y-fY'^2+f^3Y^2}{f^3Y^4},
\eq
\bq
F_V=0,
\eq
\bqn
\lb{HS}
{\cal{H}}_S&=&\frac{4}{f^7Y^8}\times\nb\\
&&[f^4Y^5A'(f'Y-2fY'')-4f''''f^2Y'Y^3+\nb\\
&&36f''f'Y'Y^3-16f''fY''Y^3+28f''f^2{Y'}^2Y^3-\nb\\
&&48{f'}^3Y'Y^3+48{f'}^2fY''Y^3-84{f'}^2f{Y'}^2Y^2-\nb\\
&&20f'f^2Y'''Y^3+100f'Y''Y'Y^2-\nb\\
&&80f'f^2{Y'}^3Y+4f^3Y''''Y^3-84{f'}^2f{Y'}^2Y^2-\nb\\
&&28f^3Y'''-16f^3{Y''}^2Y^2+88f^3Y''{Y'}^2Y+\nb\\
&&4f^5Y''Y^3-40f^3Y'^4-12f^5{Y'}^2Y^2],
\eqn
\bqn
\lb{H}
H&=&8\pi J^t =\frac{1}{f^7Y^8}\times\nb\\
&&[4f^4Y^5A'(f'Y-fY')-16f'''f^2Y'Y^3+\nb\\
&&144f''f'fY'Y^3-64f''f^2Y''Y^3+122f''f^2{Y'}^2Y^2-\nb\\
&&192f'^3Y'Y^3+192f'^2fY''Y^3-336f'^2f{Y'}^2Y^2-\nb\\
&&80f'f^2Y'''Y^3+400f'f^2Y''Y'Y^2-320f'f^2{Y'}^3Y+\nb\\
&&4f'f^4Y'Y^5\gamma_1 +16f^4Y''''Y^3-112f^3Y'''Y'Y^2-\nb\\
&&64f^4{Y''}^2Y^2+352f^3Y''{Y'}^2Y-4f^5Y''Y^4\gamma_1 +\nb\\
&&16f^5{Y'}^2Y^3-160f^3{Y'}^4+2f^5Y'Y^4\gamma_1 -\nb\\
&&48f^5{Y'}^2Y^2-9\lambda f^7\dot Y^2Y^6+3f^7\dot Y Y^6+\nb\\
&&2f^7 Y^6\gamma_1],
\eqn
\bq
\lb{Jr}
J_r=\frac{(1-3\lambda)(\dot Y' Y- \dot Y Y')}{Y^2}
\eq
\bq
\lb{JA}
J_A=\frac{2}{f^3Y^4}[2f'Y'Y-2fY''Y-f{Y'}^2+ff^3 Y^2],
\eq
\bqn
\lb{Jphi}
J_\varphi&=&\frac{1}{f^3Y^5}\times\nb\\
&&[3(1-\lambda)(f'\dot Y'Y^2+f\dot Y Y'Y-f\dot Y''Y^2)+\nb\\
&&fY''\dot Y Y+\nb\\
&&3\lambda(f'Y'\dot Y Y -fY'' \dot Y Y+fY'^2\dot Y-3f^3\dot Y Y^4)-\nb\\
&&f'Y'\dot Y Y+fY''\dot YY-2fY'^2 \dot Y+f^3\dot Y Y^2]. 
\eqn

From the dynamical equations (\ref{dyn}) we have
\bq
D^{rr}=8\pi \tau^{rr}=\frac{d^{rr}}{2f^4Y^6},
\eq
where
\bqn
\lb{drr}
d^{rr}&=&2[(A-\gamma_1)Y'^2+(A+\gamma_1)f^2Y^2]-4A'Y'Y+\nb\\
&&(1-3\lambda)(2f^2\ddot Y Y^3+\nb\\
&&f^2\dot Y^2Y^2),
\eqn
\bq
D^{\theta\theta}=8\pi \tau^{\theta\theta}=\frac{d^{\theta\theta}}{2f^3Y^6},
\eq
\bqn
\lb{dOO}
d^{\theta\theta}&=&-2A''fY^2+2A'f'Y^2+2(A+\gamma_1)\times\nb\\
&&(f'Y'Y-fY''Y+fY'^2)+(1-3\lambda)\times\nb\\
&&(2f^3\dot YY^3+f^3\dot Y^2Y^2),
\eqn
and
\bq
D^{\phi\phi}=8\pi \tau^{\phi\phi}=\sin(\theta)^2 D^{\theta\theta}.
\eq

For a spacetime filled by a dust fluid without pressure we have that 
\bq
\lb{Jr0}
J_r=0.
\eq
Using the equation (\ref{Jr}), combined with equation (\ref{Jr0}), we have that
\bq
Y(r,t)=y_1(r)y_2(t).
\eq
Taking the above condition and imposing that $J_A=0$, from equation (\ref{JA}), we obtain
\bqn
\label{Yrt}
&&2f' y_1' y_1 y_2^2-2f y_1'' y_1 y_2^2-f(y_1' y_2)^2+f^3(y_1 y_2)^2=0.\nb\\
\eqn
Solving the above equation for $f(r)$ we have that
\bq
\lb{fr}
f(r)=\frac{(y_1')^2 y_2^2}{\sqrt{2\int \frac{(y_1')^2 y_2^2}{y_1} dr +c_1}},
\eq
where $c_1$ is a constant of integration.

Since $f$ depends only on the coordinate $r$, then there is
always a combination of the functions $y_1$ and $y_2$
in the equation (\ref{fr}) that probably transforms
its right side independent of the coordinate $t$.
Substituting now, the function $f(r)$ into $J_\varphi=0$, from equation (\ref{Jphi}), we get that
\bq
y_2(t)=c_2,
\eq
where $c_2$ is an arbitrary constant.

This last equation shows that the function $Y$ does not depend on $t$, i.e., $Y(r,t)=Y(r)$.
Thus, we can rewrite the equation (\ref{fr}) as 
\bq
\lb{fr1}
f(r)=\frac{(Y')^2}{\sqrt{2\int \frac{(Y')^2 }{Y} dr +c_1}},
\eq
where we have renamed the arbritrary constant $c_1$ as $c_1\equiv c_1/c_2^2$.

From the equation (\ref{fr1}) we can conclude that the
spacetime is stationary, i.e., there is not collapse
for this kind of fluid in the IR limit of HLT.
This result is in contrast with GRT, where it is
well known that there is collapse for a pressureless fluid, possibly forming at the end a black hole.

 Let us now show that the density energy is stationary as well,
supporting the conclusion of the existence of a 
stationary spacetime.

Since $\tau^{rr}=0$ and $Y(r,t)=Y(r)$ we can write from equation (\ref{drr}) that
\bq
2[(A-\gamma_1)Y'^2+(A+\gamma_1)f^2 Y^2]-4A'Y'Y=0.
\eq
From this equation we can derive $A'$ obtaining $A''$.

Since $\tau^{\theta\theta}=0$ and substituting the
 expressions for $A'$ and $A''$ in terms of $A$ into
 equation (\ref{dOO}), we can show that
\bq
A(r,t)=A(r).
\eq

Using the equations (\ref{Jt}) and (\ref{H}) and
taking into account that $A(r,t)=A(r)$, we can show
that
\bq
\rho(r,t)=\rho(r).
\eq
This is an expected result since the metric is
independent of the coordinate $t$.  Thus, we have proved that the spacetime is stationary.

 However, it had already been shown in
the reference \cite{Lin2014} that the projectable version of HLT excludes the minimally coupling with matter. Therefore
, in the case of dust fluid, for which the metric is projectable, it was confirmed again
 in this Section, as our results have led to a static metric with a fluid
 density of dust $\rho(r,t) = \rho(r)$, contradicting the behavior of a such model
 with the same fluid in GRT.

\subsection{Shear-free Dust Solution with the Non-Minimum Coupling in the Infrared Limit}

Let us now analyze the case where it exists a non-minimum coupling with matter
\cite{Lin2014}.  In this case we have the following conditions,
\bq
\gamma_1=-1,
\eq
\bq
a_1=1,
\eq
\bq
a_2=0,
\eq
\bq
F=1-A,
\eq
\bq
\Omega=1,
\eq
\bq
\tilde N=1-A,
\eq
\bq
\tilde N^i=N^i,
\eq
\bq
\tilde g_{ij}=g_{ij},
\eq
\bq
{\cal A}=0,
\eq
\bq
\sigma=A.
\eq

With these new parameters we have found that
\bq
\lb{Jt1}
J^t=-2\rho(1-A),\nb
\eq
\bq
J^i=0,\nb
\eq
\bq
\tau^{ij}=0,
\eq
\bq
\lb{JA1}
J_A=-2\rho,\nb
\eq
\bq
J_{\varphi}=\frac{1}{Y}(Y \dot \rho+3 \dot Y \rho).\nb
\eq
Thus,
\bqn
H=8\pi J^t&=&\frac{1}{f^3Y^4}[(1-3\lambda)(3\dot Y^2 f^3Y^2)-4f'Y'fY+\nb\\
&&4Y''fY-2f^2Y'^2-2f^3Y^2],
\eqn
\bq
\lb{Jr1}
J_r=\frac{1}{f^2Y^4}(1-3\lambda)(\dot Y'-\dot Y Y'),
\eq
\bq
\lb{JA2}
J_A=\frac{2}{f^3Y^4}(2f'Y'Y-2fY''Y+fY'^2+f^3Y^2),
\eq
\bqn
\lb{Jphi1}
J_{\varphi}&=&\frac{1}{f^3Y^5}[3(1-\lambda)(f'\dot Y'Y^2+f\dot Y'Y'Y-f\dot Y''Y^2)+\nb\\
&&fY''\dot YY+(1-3\lambda)(fY''\dot Y Y-f'Y'\dot Y Y)+\nb\\
&&(3\lambda-2)fY'^2\dot Y+f^3\dot Y Y^2].
\eqn

From the dynamical equations we get
\bq
\lb{Drr1}
D^{rr}=8\pi \tau^{rr} =\frac{d^{rr}}{2f^4Y^6},
\eq
where
\bqn
d^{rr}&=&2[(1-A)(Y'^2-f^2Y^2)]-4A'Y'Y+\nb\\
&&(1-3\lambda)(2f^2\ddot Y Y^3+f^2\dot Y^2Y^2),
\eqn
\bq
\lb{D001}
D^{\theta\theta}=8\pi \tau^{\theta\theta} =\frac{d^{\theta\theta}}{2f^3Y^6},
\eq
where
\bqn
d^{\theta\theta}&=&2A'f'Y^2-2A''fY^2+2(1-A)\times\nb\\
&&(f'Y'Y+fY''Y-fY'^2)+\nb\\
&&(1-3\lambda)(2f^3\ddot Y^3+f^3\dot Y^2Y^2),
\eqn
and
\bq
D^{\phi\phi}=8\pi \tau^{\phi\phi}=D^{\theta\theta}\sin^2 \theta.
\eq

Therefore, for a spacetime filled by a fluid of dust with zero pressure and
shear-free, we have $J_r=0$, as in Section IV. Thus, solving equation
(\ref{Jr1}) for the metric function $Y(r,t)$, we have found that the solution admits separation of variables , i.e.,
\bq
Y(r,t)=y_1(r)y_2(t).
\eq

From equations (\ref{Jt1}) and (\ref{JA1}), we have that
\bq
A=1-\frac{J^t}{J_A},
\eq
thus
\bqn
\lb{A1}
A&&=\left[3(3\lambda-1)y_1^4\dot y_2 f^3+4y_1^2f^3-8y_1''y_1f+4y_1\,'^2f+\right. \nb\\
&&\left. 8f'y_1\,'y_1\right]\left[2(f^3y_1^2+2f'y_1\,'y_1-2y_1''fy_1+y_1\,'^2f)\right]^{-1}.\nb\\
\eqn

Then, we derive the above equation once and twice in relation to $r$, in order to find the
expressions for $A'$ and $A''$. With these results, we can use to the dynamic equations
(\ref{Drr1}) and (\ref{D001}) , which are identically zero. Although we know the expression for $A$,
the resulting equations are very difficult to be solved individually. However, we can notice that
\bqn
\lb{fdrr}
fd^{rr}&=&2[f(1-A)(Y'^2-f^2Y^2)]-4fA'Y'Y+\nb\\
&&(1-3\lambda)(2f^3\ddot YY^3+f^3\dot Y^2Y^2)=0.
\eqn
Thus, we can write that
\bqn
&&4fA'Y'Y-2[f(1-A)(Y'^2-f^2Y^2)]=\nb\\
&&(1-3\lambda)(2f^3\ddot YY^3+f^3\dot Y^2Y^2),
\eqn
where the right hand of this equation can be substituted in equation (\ref{fdrr}), giving
\bqn 
&&2A'f'Y^2-2A''fY^2+2(1-A)(fY'Y+fY''Y-\nb\\
&&fY'^2+4fA'Y'Y-2[f(1-A)(Y'^2-f^2Y^2)]=0.\nb\\
\eqn

Again, as in previous Section, solving this last equation
in relation to $y_1(r)$ and $y_2(t)$, we get two possible solutions
\bq
Y(r,t)=0,
\eq
or
\bq
Y(r,t)=y_1(r)(c_1t+c_2),
\eq
where $c_1$ and $c_2$ are arbitrary constants.

Since the first solution $Y(r,t)=0$ is not possible,
the only solution is the second one.  Substituting this solution into the equation (\ref{A1}) we obtain that $A(r,t)=A(r)$.

Now, solving the equation for $J_{\varphi}$ and using the equations (\ref{Jt1}) to  (\ref{Jphi1})
we have that
\bq
\lb{rhort}
\rho(r,t)=\frac{c_1\rho_1(r)+F(r)}{(c_1t+c_2)^3},
\eq
where the function $\rho_1(r)$ and the arbitrary function $F(r)$ are dependent only to $r$.

Using the equation (\ref{JA1}) for $J_A$ we get that
\bq
\rho=-\frac{2f'Y'Y-2fY''Y+fY'^2+f^3Y^2}{f^3Y^4}.
\eq

Using the equations (\ref{JA2}) and (\ref{rhort}) we get three possible solutions:
\begin{enumerate}
\item $f(r)=0$, which is physically impossible,
\item $c_1=0$, which gives the same results without the minimally coupling,
\item $F(r)=0$, which allows the density $\rho$ dependent of $t$, i.e., $\rho(r,t)$.
\end{enumerate}
With the third condition, we obtain that
\bq
\lb{rhort1}
\rho(r,t)=\frac{c_1\rho_1(r)t}{(c_1t+c_2)^3}.
\eq

We can see that the non-minimum coupling produces interesting results.
It is still necessary analyze the temporal behavior of the energy density, in order to verify if this solution admits a gravitational collapse process, as expected in GRT. Thus, deriving the energy density given in the equation (\ref{rhort1}), we get
\bq
\dot \rho(r,t)=\frac{c_1\rho_1(r)(c_2-2c_1t)}{(c_1t+c_2)^4}.
\eq

Analyzing $\rho(r,t)$ at $t=0$, we have
\bq
\dot \rho(r,t=0)=\frac{c_1\rho_1(r)}{c_2^3}.
\eq
Since the energy density as well as its temporal rate of change should be
always positive, in order to insure a physically acceptable collapsing fluid, we have to
impose some conditions on the constants $c_1$, $c_2$ and on the function $\rho_1(r)$.

\begin{figure}
\vspace{.2in}
\centerline{\psfig{figure=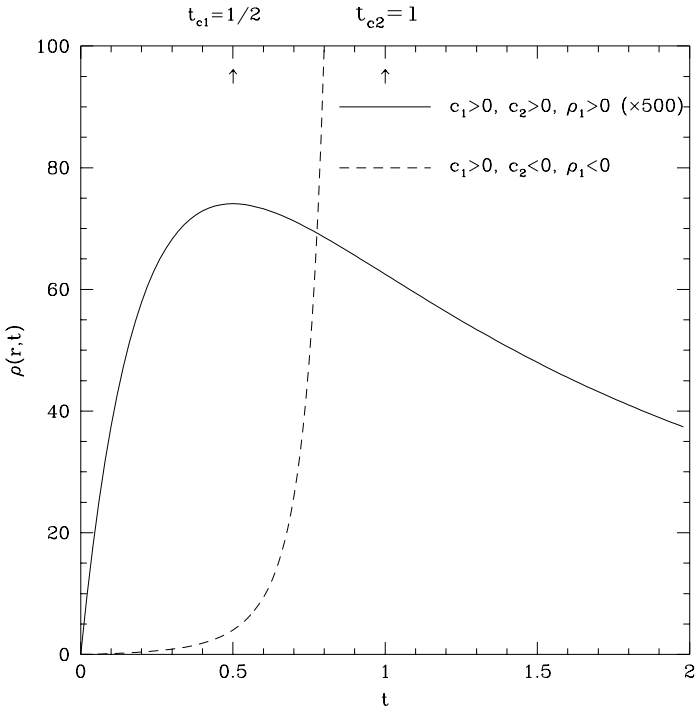,width=3.8truein,height=3.8truein}\hskip .1in}\caption{Temporal behavior of the density $\rho(r,t)$, given by equation (\ref{rhort1}). For $\rho_1>0$ we have used the values $\rho_1=1$, $c_1=1$ and $c_2=1$, where $t=t_{c1}$ represents the time of the inversion of sign of $\dot \rho$. For $\rho_1<0$ we have used the values $\rho_1=-1$, $c_1=1$ and $c_2=-1$, where $t=t_{c2}$ denotes the time of divergence of $\dot \rho$.}
\label{fig1}
\end{figure}

Let us study two cases: Case (1) $\rho_1(r)>0$ and Case (2) $\rho_1(r)<0$ (see Figure \ref{fig1}).
\begin{enumerate}
\item
Since $\rho\ge 0$, them $c_1$ and $c_2$ must have the same signs. Besides, when $\dot \rho>0$ then $0\le t<t_{c1}$ and $\dot \rho<0$ then $t>t_{c1}$, where $t=t_{c1}=|c_1|/(2|c_2|)$ represents the time of the inversion of sign of $\dot \rho$. Thus, this case describes a situation of an initial contraction followed by an
 expansion, without the formation of a singularity.
\item
Again, since $\rho\ge 0$, thus $c_1$ and $c_2$ must have the opposite signs. Besides, when $\dot \rho>0$ then $0\le t<t_{c2}$ and $\rho \rightarrow \infty$ at $t=t_{c2}=|c_2|/|c_1|$. Thus, this case describes a typical gravitational collapse situation, which is
 consistent with the results of GRT.
\end{enumerate}

\section{Conclusion}

In this present work, using again the results of Lin et al. (2014) \cite{Lin2014}
we have studied the spherically 
symmetric spacetime filled by a dust fluid,  in general covariant theory ($U(1)$ extension) of
HLT with the minimally coupling \cite{Lin2014}, in the
PPN approximation in the infrared 
limit.  We have analyzed if a solution like this one can be described in the general
covariant HLT of gravity \cite{ZWWS,ZSWW}.  

Although we do not have a realistic model, we can describe the
gravitational collapse as we can see in GRT. Besides, confirming what was found in the reference \cite{Lin2014}, the projectable HLT
excludes the minimally coupling, and it does not reproduce the well known results in GRT.

However, when using non-minimum coupling with matter, we have shown that the solution
admits a process of gravitational collapse, leaving a singularity at the end.
Note that we have also found a second possible solution, representing a bounce behavior that is 
not expected in GRT.

\begin{acknowledgements}

The financial assistance from FAPERJ/UERJ\\   (MFAS) are gratefully acknowledged.
The author (RC) acknowledges the financial support from FAPERJ 
(no. E-26/171.754/2000, E-26/171.533/2002, E-26/170.951/2006, E-26/110.432/2009 and\\ E-26/111.714/2010). The authors (RC and MFAS) also acknowledge the 
financial support from Conselho Nacional de Desenvolvimento Cient\'ifico e
Tecnol\'ogico - CNPq - Brazil (no. 450572/2009-9, 301973/2009-1 and 
477268\-/2010-2). The author (MFAS) also acknowledges the financial support 
from Financiadora de Estudos e Projetos - FINEP - Brazil (Ref. 2399/03).
The authors (OG and MFAS) also thank the financial support from CAPES/Science without Borders (no. A 045/2013).
We also would like to thank Dr. Anzhong Wang for helpful discussions and
comments about this work.

\end{acknowledgements}

\section{Appendix A: Definition of $F^{ij}$, $F_S^{ij}$, $F^{ij}_a$, $F^{ij}_\varphi$ and $F_S^{ij}$}

The quantities $F^{ij}$, $F_S^{ij}$, $F^{ij}_a$, $F^{ij}_\varphi$ and 
$F_S^{ij}$ are given by

\bqn
F^{ij}&=&\frac{1}{\sqrt{g}N}\frac{\delta (-\sqrt{g}N {\cal{L}}_V^R)}{\delta g_{ij}}\nb\\
&=& \sum_{s=0}\hat{\gamma}_s\zeta^{n_s}(F_s)^{ij},nb\\
F_S^{ij}&=&-\sigma \left(\sigma_1a^ia^j+\sigma_2a^{ij}\right)\nb\\
&&+\frac{a_S}{2}\left[(\nabla^i\varphi)(\nabla^j\varphi)+2\frac{N^{(i}\nabla^{j)}\varphi}{N}\right] \nb\\                              &&+\frac{\sigma_2}{N}\nabla^{(i}[a^{j)}(A-{\cal
                    A})]-g^{ij}\frac{\sigma_2}{2N}\nabla^{k}[a_k(A-{\cal
                    A})],\nb\\
F^{ij}_a&=&\frac{1}{\sqrt{g}N}\frac{\delta (-\sqrt{g}N {\cal{L}}_V^a)}{\delta g_{ij}}\nb\\
& =& \sum_{s=0}\beta_s\zeta^{m_s}(F_s^a)^{ij},\\
F^{ij}_\varphi&=&\frac{1}{\sqrt{g}N}\frac{\delta (-\sqrt{g}N {\cal{L}}_V^\varphi)}{\delta g_{ij}}\nb\\
&=& \sum_{s=0}\mu_s(F_s^\varphi)^{ij},\nb\\
F_S^{ij}&=&-\sigma \left(\sigma_1a^ia^j+\sigma_2a^{ij}\right)\nb\\
                &&+\frac{a_S}{2}\left[(\nabla^i\varphi)(\nabla^j\varphi)+2\frac{N^{(i}\nabla^{j)}\varphi}{N}\right] \nb\\
                    &&+\frac{\sigma_2}{N}\nabla^{(i}[a^{j)}(A-{\cal
                    A})]-g^{ij}\frac{\sigma_2}{2N}\nabla^{k}[a_k(A-{\cal
                    A})],\nb\\
\eqn
with
\bqn
\hat{\gamma}_s &=& \left(\gamma_0, \gamma_1, \gamma_2, \gamma_3, \frac{1}{2}\gamma_5, -\frac{5}{2}\gamma_5, 3\gamma_5, \frac{3}{8}\gamma_5, \gamma_5, \frac{1}{2}\gamma_5\right), \nb\\
n_s &=& (2, 0, -2, -2, -4, -4, -4, -4, -4, -4),\nb\\
m_s&=& (0, -2,-2,-2, -2, -2, -2, -2, -4 ), \nb\\
\mu_s &=& \left(2, 1, 1, 2, \frac{4}{3}, \frac{5}{3}, \frac{2}{3}, 1-\lambda, 2-2 \lambda\right).
\eqn

Thus, $F_V,\;F_\varphi$ and $F_\lambda$ are given, respectively, by
\bqn\label{a1}
F_V &=&  \beta_0 ( 2 a_i^i + a_i a^i) - \frac{\beta_1}{\zeta^2} \Bigg[3 (a_i a^i)^2 + 4 \nabla_i (a_k a^k a^i)\Bigg]\nb\\
    &&  +\frac{\beta_2}{\zeta^2}\Bigg[ (a_i^i)^2 + \frac{2}{N} \nabla^2 (N a_k^k)\Bigg]\nb\\
    && + \frac{\beta_3}{\zeta^2}\Bigg[ - (a_i a^i) a_j^j - 2 \nabla_i (a_j^j a^i) + \frac{1}{N} \nabla^2 (N a_i a^i)\Bigg]\nb\\
    &&+ \frac{\beta_4}{\zeta^2}\Bigg[ a_{ij} a^{ij} + \frac{2}{N} \nabla_j \nabla_i (N a^{ij})\Bigg]\nb\\
      && + \frac{\beta_5}{\zeta^2}\Bigg[- R (a_i a^i) - 2 \nabla_i (R a^i)\Bigg]\nb\\
      &&+ \frac{\beta_6}{\zeta^2}\Bigg[- a_i a_j R^{ij} - \nabla_i (a_j R^{ij})-\nabla_j (a_i R^{ij})\Bigg]\nb\\
      && +  \frac{\beta_7}{\zeta^2}\Bigg[ R a^i_i + \frac{1}{N} \nabla^2 (NR)\Bigg]\nb\\
      &&+ \frac{\beta_8}{\zeta^4}\Bigg[(\Delta a^i)^2 - \frac{2}{N} \nabla^i [\Delta (N \Delta a_i)]\Bigg],
\eqn
\bqn\label{a2}
F_\varphi &=& -  {\cal{G}}^{ij}\nabla_i \varphi \nabla_j \varphi, - \frac{2}{N} \hat{{\cal{G}}}^{ijkl} \nabla_l (N K_{ij} \nabla_k \varphi),\nb\\
        &&  - \frac{4}{3}\Bigg[ \hat{{\cal{G}}}^{ijkl} \nabla_l (\nabla_k \varphi \nabla_i \nabla_j \varphi)\Bigg]\nb\\
         &&+ \frac{5}{3}\Bigg[ -  \hat{{\cal{G}}}^{ijkl} [(a_i \nabla_j \varphi) (a_k \nabla_l \varphi)+\nabla_i(a_k\nabla_j\varphi\nabla_l\varphi)\nb\\
         && +\nabla_k(a_i\nabla_j\varphi\nabla_l\varphi)]\Bigg]\nb\\
         &&+ \frac{2}{3}\Bigg[ \hat{{\cal{G}}}^{ijkl}[a_{ik} \nabla_j \varphi \nabla_l \varphi + \frac{1}{N} \nabla_i\nabla_k (N\nabla_j\varphi\nabla_l\varphi)]\Bigg],\nb\\
\eqn
\bqn\lb{a3a}
F_\lambda &=& (1-\lambda) \Bigg\{(\nabla^2 \varphi + a_i \nabla^i \varphi)^2 - \frac{2}{N} \nabla_i (NK\nabla^i \varphi)\nb\\
           && - \frac{2}{N} \nabla_i [N (\nabla^2 \varphi + a_i \nabla^i \varphi) \nabla^i \varphi]\Bigg\}\label{a3}.
\eqn

$\left(F_n\right)_{ij}$, $\left(F^{a}_{s}\right)_{ij}$ and $\left(F^{\varphi}_{q}\right)_{ij}$, defined in equation (\ref{tauij}),  are given, respectively, by
\bqn
(F_0)_{ij} &=& -\frac{1}{2}g_{ij},\nb\\
(F_1)_{ij} &=& R_{ij}-\frac{1}{2}Rg_{ij}+\frac{1}{N}(g_{ij}\nabla^2 N-\nabla_j\nabla_i N),\nb\\
(F_2)_{ij} &=& -\frac{1}{2}g_{ij}R^2+2RR_{ij}\nb\\
             &&  +\frac{2}{N}\left[g_{ij}\nabla^2(NR)-\nabla_j\nabla_i(NR)\right],\nb\\
(F_3)_{ij} &=& -\frac{1}{2}g_{ij}R_{mn}R^{mn}+2R_{ik}R^k_{j}\nb\\
             &&  +\frac{1}{N}\Big[- 2\nabla_k\nabla_{(i}(NR_{j)}^k)\nb\\
             &&  +\nabla^2(NR_{ij})+g_{ij}\nabla_m\nabla_n(NR^{mn})\Big], \nb\\
(F_4)_{ij} &=&  -\frac{1}{2}g_{ij}R^3+3R^2R_{ij}\nb\\
             &&   +\frac{3}{N}\Big(g_{ij}\nabla^2-\nabla_j\nabla_i\Big)(NR^2),\nb\\
(F_5)_{ij} &=& -\frac{1}{2}g_{ij}RR_{mn}R^{mn}\nb\\
             &&+R_{ij}R_{mn}R^{mn}+2RR_{ik}R^k_{j}\nb\\
             &&  +\frac{1}{N}\Big[g_{ij}\nabla^2(NR_{mn}R^{mn})\nb\\
             &&-\nabla_j\nabla_i(NR_{mn}R^{mn})\nb\\
             &&  +\nabla^2(NRR_{ij})+g_{ij}\nabla_m\nabla_n(NRR^{mn})\nb\\
             &&  -2\nabla_m\nabla_{(i}(R^m_{j)}NR)\Big], \nb\\
(F_6)_{ij} &=& -\frac{1}{2}g_{ij}R^m_nR^n_lR^l_m+3R^{mn}R_{mi}R_{nj}\nb\\
             &&  +\frac{3}{2N}\Big[g_{ij}\nabla_m\nabla_n(NR^m_aR^{na}) \nb\\
             &&+ \nabla^2(NR_{mi}R^m_j)
              -2\nabla_m\nabla_{(i}(NR_{j)n}R^{mn})\Big], \nb\\
(F_7)_{ij} &=& -\frac{1}{2}g_{ij}R\nabla^2R+R_{ij}\nabla^2R+R\nabla_i\nabla_j R\nb\\
             &&  +\frac{1}{N}\Big[g_{ij}\nabla^2(N\nabla^2R)-\nabla_j\nabla_i(N\nabla^2R)\nb\\
             &&+R_{ij}\nabla^2(NR)
              +g_{ij}\nabla^4(NR)-\nabla_j\nabla_i (\nabla^2 (NR))\nb\\
             &&  - \nabla_{(j}(NR\nabla_{i)}R)+\frac{1}{2}g_{ij}\nabla_k(NR\nabla^kR)\Big], \nb\\
(F_8)_{ij} &=& -\frac{1}{2}g_{ij}(\nabla_mR_{nl})^2 + 2 \nabla^mR^n_i\nabla_mR_{nj}\nb\\
             &&  +\nabla_iR^{mn}\nabla_jR_{mn}+\frac{1}{N}\Big[2 \nabla_n\nabla_{(i}\nabla_m(N\nabla^mR^n_{j)})\nb\\
             &&  -\nabla^2\nabla_m(N\nabla^mR_{ij})-g_{ij}\nabla_n\nabla_p\nabla_m(N\nabla^mR^{np})\nb\\
             &&  -2\nabla_m(NR_{l(i}\nabla^mR^l_{j)})-2\nabla_n(NR_{l(i}\nabla_{j)}R^{nl})\nb\\
             &&  +2\nabla_k(NR^k_l\nabla_{(i}R^l_{j)})\Big], \nb\\
(F_9)_{ij} &=& -\frac{1}{2} g_{ij} a_k G^k+\frac{1}{2} \Big[a^k R_{k(j} \nabla_{i)} R + a_{(i} R_{j)k} \nabla^k R\Big]\nb\\
           &&-a_kR_{mi}\nabla_jR^{mk}-a^kR_{n(i}\nabla^nR_{j)k}\nb \\
           &&-\frac{1}{2}\Big[a_iR^{km}\nabla_mR_{kj}+a_jR^{km}\nabla_mR_{ki}\Big]\nb\\
           &&-\frac{3}{8}a_{(i}R\nabla_{j)}R+\frac{3}{8}\Bigg\{R\nabla_k(Na^k)R_{ij}\nb\\
           &&+g_{ij}\nabla^2\Big[R\nabla_k(Na^k)\Big]-\nabla_i\nabla_j\Big[R\nabla_k(Na^k)\Big]\Bigg\}\nb\\
           &&+\frac{1}{4N} \Bigg\{- \frac{1}{2}\nabla^m \Big[\nabla_{(i} Na_{j)}\nabla_m R+\nabla_{(i}(\nabla_{j)}R) Na_m\Big]\nb\\
           &&+\nabla^2 (N a_{(i}\nabla_{j)}R)+g_{ij} \nabla^m\nabla^n (Na_m\nabla_nR)\nb\\
           && +\nabla^m\Big[\nabla_{(i} (\nabla_{j)} R^k_m) Na_k+\nabla_{(i}(\nabla_m R^k_{j)})Na_k\Big]\nb\\
           &&-2\nabla^2(Na_k\nabla_{(i} R^k_{j)})-2g_{ij}\nabla^m\nabla^n(Na_k\nabla_{(n}R_{m)}^k)\nb\\
           &&- \nabla^m \Big[\nabla_i\nabla_p(Na_jR_m^p+Na_mR_j^p)\nb\\
           &&+\nabla_j\nabla_p(Na_iR_m^p+Na_mR_i^p)\Big]\nb\\
           &&+2\nabla^2\nabla_p(Na_{(i}R_{j)}^p)\nb\\
           && +2g_{ij}\nabla^m\nabla^n \nabla^p(Na_{(n}R_{m)p})\Bigg\}, \nb\\
\eqn

\bqn
(F_0^a)_{ij} &=&  -\frac{1}{2} g_{ij} a^k a_k +a_i a_j, \nb\\
(F_1^a)_{ij} &=&  -\frac{1}{2} g_{ij} (a_k a^k)^2 + 2 (a_k a^k) a_i a_j,\nb\\
(F_2^a)_{ij} &=&  -\frac{1}{2} g_{ij} (a_k^k)^2 + 2 a_k^k a_{ij}\nb\\
             &&   - \frac{1}{N} \Big[2 \nabla_{(i} (N a_{j)} a_k^k) - g_{ij} \nabla_\alpha (a_\alpha N a_k^k)\Big],\nb\\
(F_3^a)_{ij} &=&   -\frac{1}{2} g_{ij} (a_k a^k) a_\beta^\beta + a^k_k a_ia_j + a_k a^k a_{ij}\nb\\
             &&   - \frac{1}{N} \Big[ \nabla_{(i} (N a_{j)} a_k a^k) - \frac{1}{2} g_{ij} \nabla_\alpha (a_\alpha N a_ka^k)\Big],\nb\\
(F_4^a)_{ij} &=&  - \frac{1}{2} g_{ij} a^{mn} a_{mn} + 2a^k_i a_{kj} \nb\\
             &&   - \frac{1}{N} \Big[\nabla^k (2 N a_{(i} a_{j)k} - N a_{ij} a_k)\Big], \nb\\
(F_5^a)_{ij} &=&  -\frac{1}{2} g_{ij} (a_k a^k ) R + a_i a_j R + a^k a_k R_{ij} \nb\\
              &&  + \frac{1}{N} \Big[ g_{ij} \nabla^2 (N a_k a^k) - \nabla_i \nabla_j (N a_k a^k)\Big], \nb\\ %\\
(F_6^a)_{ij} &=&   -\frac{1}{2} g_{ij} a_m a_n R^{mn} +2 a^m R_{m (i} a_{j)} \nb\\
              &&  - \frac{1}{2N} \Big[ 2 \nabla^k \nabla_{(i} (a_{j)} N a_k) - \nabla^2 (N a_i a_j) \nb\\
               && - g_{ij} \nabla^m \nabla^n (N a_m a_n)\Big], \nb\\ %\
(F_7^a)_{ij} &=&  -\frac{1}{2} g_{ij} R a_k^k + a_k^kR_{ij} + R a_{ij} \nb\\
             &&   + \frac{1}{N} \Big[ g_{ij} \nabla^2 (N a_k^k) - \nabla_i \nabla_j (N a_k^k) \nb\\
             &&  - \nabla_{(i} (N R a_{j)}) + \frac{1}{2} g_{ij} \nabla^k (N R a_k)\Big], \nb\\
(F_8^a)_{ij} &=&  -\frac{1}{2} g_{ij} (\Delta a_k)^2 + (\Delta a_i) (\Delta a_j) + 2 \Delta a^k \nabla_{(i} \nabla_{j)} a_k \nb\\
             &&   + \frac{1}{N} \Big[\nabla_k [a_{(i} \nabla^k (N \Delta a_{j)}) + a_{(i} \nabla_{j)} (N \Delta a^k)\nb\\
             &&   - a^k \nabla_{(i} (N \Delta a_{j)}) + g_{ij} N a^{\beta k} \Delta a_\beta  - N a_{ij} \Delta a^k ]\nb\\
             &&  -  2 \nabla_{(i} (N a_{j)k} \Delta a^k)\Big],
\eqn

\bqn
(F_1^\varphi)_{ij} &=&     -\frac{1}{2} g_{ij} \varphi {\cal{G}}^{mn}K_{mn}\nb\\
                   &&  + \frac{1}{2\sqrt{g} N}  \partial_t (\sqrt{g} \varphi {\cal{G}}_{ij}) -
                                                                                2 \varphi K_{(i}^\nu R_{j) \nu} \nb\\
                   &&       + \frac{1}{2} \varphi (K R_{ij} + K_{ij} R ) \nb\\
                   &&      + \frac{1}{2N} \bigg\{{\cal{G}}_{ij} \nabla^k (\varphi N_k) - 2 {\cal{G}}_{k(i} \nabla^k (N_{j)} \varphi)\nb\\
                   &&      +  g_{ij}\nabla^2 (N \varphi K) - \nabla_i \nabla_j (N \varphi K) \nb\\
                   &&+ 2  \nabla^k \nabla_{(i} (K_{j)k} \varphi N),\nb\\
                   &&      - \nabla^2 (N \varphi K_{ij}) - g_{ij} \nabla^\alpha \nabla^\beta (N \varphi K_{\alpha \beta})\bigg\}, \nb\\
(F_2^\varphi)_{ij} &=&   - \frac{1}{2}g_{ij} \varphi {\cal{G}}^{mn} \nabla_m\nabla_n \varphi \nb\\
                   &&     - 2 \varphi \nabla_{(i} \nabla^k R_{j)k} + \frac{1}{2} \varphi R \nabla_i \nabla_j \varphi \nb\\
                   &&     - \frac{1}{N} \bigg\{- \frac{1}{2} (R_{ij} + g_{ij} \nabla^2 - \nabla_i \nabla_j )(N \varphi \nabla^2 \varphi)\nb\\
                   &&      - \nabla_k \nabla_{(i} (N \varphi \nabla^k \nabla_{j)} \varphi) + \frac{1}{2}\nabla^2 (N \varphi \nabla_i \nabla_j \varphi) \nb\\
                   &&      + \frac{g_{ij}}{2} \nabla^\alpha \nabla^\beta ( N \varphi \nabla_\alpha \nabla_\beta \varphi)\nb\\
                   &&      - {\cal{G}}_{k (i} \nabla^k (N \varphi \nabla_{j)}\varphi ) + \frac{1}{2} {\cal{G}}_{ij} \nabla^k (N \varphi \nabla_k \varphi)\bigg\},\nb\\
(F_3^\varphi)_{ij} &=&    - \frac{1}{2}g_{ij} \varphi {\cal{G}}^{mn} a_m\nabla_n \varphi \nb\\
                   &&    -\varphi ( a_{(i} R_{j)k} \nabla^k \varphi + a^k R_{k(i} \nabla_{j)} \varphi)\nb\\
                   &&      + \frac{1}{2} R \varphi a_{(i} \nabla_{j)} \varphi \nb\\
                   &&      - \frac{1}{N}\bigg\{ - \frac{1}{2} (R_{ij} + g_{ij} \nabla^2 - \nabla_i \nabla_j ) (N \varphi a^k \nabla_k \varphi)\nb\\
                   &&     - \frac{1}{2} \nabla^k \Big[  \nabla_{(i} (\nabla_{j)} \varphi N \varphi)+ \nabla_{(i} (a_{j)} \varphi N \nabla_k \varphi) \Big] \nb\\
                   &&     + \frac{1}{2}\nabla^2 (N \varphi a_{(i} \nabla_{j)} \varphi) \nb\\
                   &&+ \frac{g_{ij}}{2} \nabla^\alpha \nabla^\beta (N \varphi a_\alpha \nabla_\beta \varphi)\bigg\}, \nb\\ 
(F_4^\varphi)_{ij} &=&    - \frac{1}{2}g_{ij} \hat{{\cal{G}}}^{mnkl}K_{mn} a_{(k}\nabla_{l)}\varphi \nb\\
                   &&    + \frac{1}{2 \sqrt{g} N} \partial_t [\sqrt{g} {\cal{G}}_{ij}^{\;\;k l} a_{(l} \nabla_{k)} \varphi] \nb\\
                   &&      + \frac{1}{2N} \nabla^\alpha \Big[  a_\alpha N_{(i} \nabla_{j)} \varphi +  N_{(i} a_{j)} \nabla_\alpha  \varphi \nb\\
                   &&      -  N_\alpha a_{(i} \nabla_{j)} \varphi + 2 g_{ij} N_\alpha a^k \nabla_k \varphi \Big] \nb\\
                   &&      + \frac{1}{N} \nabla_{(i} (N N_{j)} a^k \nabla_k \varphi)\nb\\
                   &&      + a^k K_{k(i} \nabla_{j)} \varphi + a_{(i} K_{j)k} \nabla^k \varphi \nb\\
                   &&      - K a_{(i} \nabla_{j)} \varphi - K_{ij} a^k \nabla_k \varphi, \nb\\
(F_5^\varphi)_{ij} &=&     -\frac{1}{2} g_{ij} \hat{{\cal{G}}}^{mnkl}[a_{(k}\nabla_{l)}\varphi][\nabla_m\nabla_n\varphi]\nb\\
                    && -a_{(i} \nabla^k \nabla_{j)} \varphi \nabla_k \varphi - a_k \nabla^k \nabla_{(i} \varphi \nabla_{j)}\varphi\nb\\
                   &&     + a_{(i} \nabla_{j)} \varphi \nabla^2 \varphi + a^k \nabla_k \varphi \nabla_i\nabla_j \varphi \nb\\
                   &&     + \frac{1}{2N} \bigg\{ \nabla^k (N \varphi a_k \nabla_i \varphi \nabla_j \varphi) \nb\\
                   &&   - 2 \nabla_{(i} (N \nabla_{j)} \varphi a^k
                          \nabla_k \varphi) \nb\\
                   &&     +g_{ij} \nabla^\alpha (\nabla_\alpha \varphi a^k \nabla_k \varphi)\bigg\}, \nb\\
(F_6^\varphi)_{ij} &=&   - \frac{1}{2} g_{ij}\hat{{\cal{G}}}^{mnkl} [a_{(m}\nabla_{n)}\varphi][a_{(k}\nabla_{l)}\varphi]\nb\\
                     &&-\frac{1}{2} (a^k \nabla_i \varphi- a_i \nabla^k \varphi) (a_k \nabla_j \varphi - a_j \nabla_k \varphi), \nb\\
(F_7^\varphi)_{ij} &=&   -\frac{1}{2} g_{ij}   \hat{{\cal{G}}}^{mnkl} [\nabla_{(n}\varphi][a_{m)(k}][\nabla_{l)}\varphi]\nb\\
                    && -\frac{1}{2} a_k^k \nabla_i \varphi \nabla_j \varphi - \frac{1}{2} a_{ij} \nabla^k \varphi \nabla_k \varphi \nb\\
                     &&     +  a^k_{(i} \nabla_{j)} \varphi \nabla_k \varphi - \frac{1}{2N}\bigg \{- \nabla_{(i} (N a_{j)} \nabla_k \varphi \nabla^k \varphi) \nb\\ &&+ \nabla^k (N a_{(i} \nabla_{j)} \varphi \nabla_k \varphi)\nb\\
                   &&      + \frac{g_{ij}}{2} \nabla^k (N a_k \nabla^m \varphi \nabla_m \varphi) \nb\\
                   &&- \frac{1}{2}\nabla^k (N a_k \nabla_i \varphi \nabla_j \varphi)\bigg\}, \nb\\
                   %\\
(F_8^\varphi)_{ij} &=&    - \frac{1}{2} g_{ij} (\nabla^2 \varphi+a_k\nabla^k\varphi)^2\nb\\
                   &&    -2 (\nabla^2 \varphi + a_k \nabla^k \varphi) (\nabla_i \nabla_j \varphi + a_i \nabla_j \varphi ) \nb\\
                   &&      -\frac{1}{N} \bigg \{ - 2 \nabla_{(j} [N \nabla_{i)} \varphi (\nabla^2 \varphi + a_k \nabla^k \varphi)] \nb\\
                   &&       + g_{ij} \nabla^\alpha [N (\nabla^2 \varphi + a_k \nabla^k \varphi) \nabla_\alpha \varphi]\bigg\}, \nb\\
(F_9^\varphi)_{ij} &=&    - \frac{1}{2} g_{ij}(\nabla^2 \varphi+a_k\nabla^k\varphi)K \nb\\
                    && - (\nabla^2 \varphi + a_k \nabla^k \varphi) K_{ij} \nb\\
                   &&     - (\nabla_i \nabla_j \varphi + a_i \nabla_j \varphi ) K \nb\\
                  &&     +\frac{1}{2 \sqrt{g} N}  \partial_t [\sqrt{g} (\nabla^2 \varphi + a_k \nabla^k \varphi) g_{ij}] \nb\\
                    &&      - \frac{1}{N}\bigg\{ - \nabla_{(j} [ N_{i)} (\nabla^2 \varphi + a_k \nabla^k \varphi)] \nb\\
                    &&     + \frac{1}{2} g_{ij} \nabla_\alpha [ N_\alpha (\nabla^2 \varphi + a_k \nabla^k \varphi) ]\nb\\
                    &&    -  \nabla_{(j} (N K \nabla_{i)} \varphi) + \frac{1}{2} g_{ij} \nabla_k (N K \nabla^k \varphi)
                            \bigg\}.\nb\\
\eqn

\end{document}